\documentclass[prl,twocolumn,showpacs,preprintnumbers,amsmath,amssymb,superscriptaddress]{revtex4}

\usepackage{graphicx}
\usepackage{pifont} 

\begin{document}


\title{Quasiparticle relaxation in optically excited high-Q superconducting resonators}

\author{R. Barends}
\affiliation{Kavli Institute of NanoScience, Faculty of Applied
Sciences, Delft University of Technology, Lorentzweg 1, 2628 CJ
Delft, The Netherlands}

\author{J. J. A. Baselmans}
\affiliation{SRON Netherlands Institute for Space Research,
Sorbonnelaan 2, 3584 CA Utrecht, The Netherlands}

\author{S. J. C. Yates}
\affiliation{SRON Netherlands Institute for Space Research,
Sorbonnelaan 2, 3584 CA Utrecht, The Netherlands}

\author{J. R. Gao}
\affiliation{Kavli Institute of NanoScience, Faculty of Applied
Sciences, Delft University of Technology, Lorentzweg 1, 2628 CJ
Delft, The Netherlands}

\affiliation{SRON Netherlands Institute for Space Research,
Sorbonnelaan 2, 3584 CA Utrecht, The Netherlands}

\author{J. N. Hovenier}
\affiliation{Kavli Institute of NanoScience, Faculty of Applied
Sciences, Delft University of Technology, Lorentzweg 1, 2628 CJ
Delft, The Netherlands}

\author{T. M. Klapwijk}
\affiliation{Kavli Institute of NanoScience, Faculty of Applied
Sciences, Delft University of Technology, Lorentzweg 1, 2628 CJ
Delft, The Netherlands}

\date{\today}

\begin{abstract}
The quasiparticle relaxation time in superconducting films has been
measured as a function of temperature using the response of the
complex conductivity to photon flux. For tantalum and aluminium,
chosen for their difference in electron-phonon coupling strength, we
find that at high temperatures the relaxation time increases with
decreasing temperature, as expected for electron-phonon interaction.
At low temperatures we find in both superconducting materials a
saturation of the relaxation time, suggesting the presence of a
second relaxation channel not due to electron-phonon interaction.
\end{abstract}

\maketitle

The equilibrium state of a superconductor at finite temperatures
consists of the Cooper pair condensate and thermally excited
quasiparticles. The quasiparticle density $n_{qp}$ decreases
exponentially with decreasing temperature. These charge carriers
control the high frequency ($\omega$) response of the superconductor
through the complex conductivity $\sigma_1- i \sigma_2$. At nonzero
frequencies the real part $\sigma_1$ denotes the conductivity by
quasiparticles and the imaginary part $\sigma_2$ is due to the
superconducting condensate \cite{tinkham,mattis}. When the
superconductor is driven out of equilibrium it relaxes back to the
equilibrium state by the redistribution of quasiparticles over
energy and by recombination of quasiparticles to Cooper pairs. The
recombination is a binary reaction, quasiparticles with opposite
wavevector and spin combine, and the remaining energy is transferred
to another excitation. The latter process is usually controlled by
the material dependent electron-phonon interaction
\cite{miller,kaplan}. With decreasing temperatures the recombination
time increases exponentially reflecting the reduced availability of
quasiparticles. Here, we report relaxation time measurements in
superconducting films far below the critical temperature $T_c$. We
find strong deviations from exponentially rising behavior, which we
attribute to the emergence of an additional relaxation channel in
the superconducting films.

\begin{figure}[b!]
    \centering
    \includegraphics[width=1\linewidth]{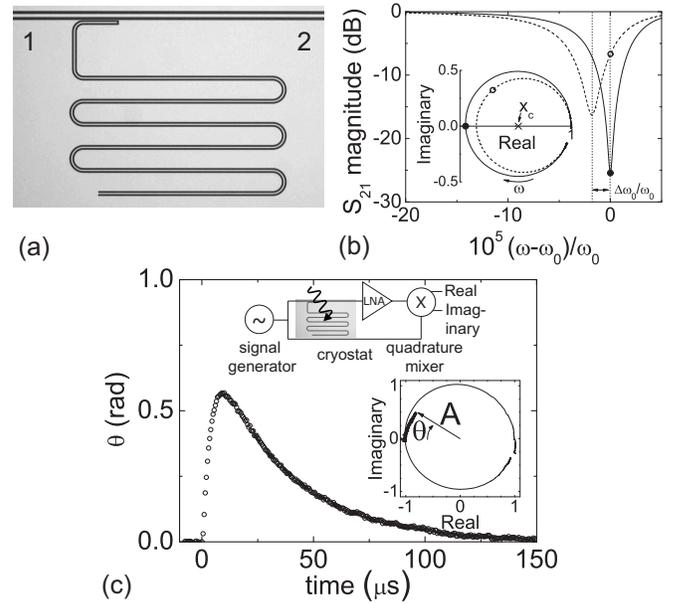}
    \caption{(a) A quarter wavelength resonator, capacitively coupled to a feedline,
    formed by the superconducting film (gray) interrupted by slits (black).
    (b) The resonator exhibits a dip in the magnitude and circle in the complex plane (inset) of the feedline
    transmission $S_{21}$. (c) The feedline transmission is converted into a phase $\theta$ and amplitude $A$ using the equilibrium resonance circle as
    reference (right inset). The response to an optical pulse of length 0.5~$\mu$s (at $t$=0) (open circles) exhibits an initial rise due to the response time (3.7~$\mu$s) of the resonator and
    subsequently follows an exponential decay (34~$\mu$s) (dashed), reflecting the restoration of equilibrium (Eq.~\ref{equation:theta}).
    The response is measured with a signal generator, low noise amplifier (LNA) and quadrature mixer (upper inset).}
    \label{figure:fig1}
\end{figure}

We have measured the time dependence of the complex conductivity of
superconducting films after applying an optical photon pulse. In
addition, the noise spectrum is measured in the presence of a
continuous photon flux \cite{wilson2001}. The superconducting film
is patterned as a planar microwave resonator. The resonator is
formed by a meandering coplanar waveguide (CPW), with the central
line 3 $\mu$m and the slits 2 $\mu$m wide, and is coupled to a
feedline, see Fig. \ref{figure:fig1}a \cite{day}. The complex
conductivity results in a kinetic inductance $L_k \propto
1/d\omega\sigma_2$, for thin films with thickness $d$, which is due
to the inertia of the Cooper pair condensate. It sets together with
the length of the central line the resonance frequency:
$\omega_0=2\pi /4l \sqrt{(L_g+L_k)C}$, with $l$ the length of a
quarterwave resonator, $L_g$ the geometric inductance and $C$ the
capacitance, both per unit length. The variation in kinetic
inductance due to photons is connected to the quasiparticle density
$n_{qp}$ by $\delta L_k/L_k~=~\frac{1}{2}\delta n_{qp}/n_{cp}$, with
$n_{cp}$ the Cooper pair density ($n_{qp} \ll n_{cp}$). Resonance
frequencies used lie between 3-6 GHz. For a quarterwave resonator at
6 GHz, the length of the meandering superconducting CPW-line is 5
mm. The resonator is capacitively coupled by placing a part parallel
to the feedline.

The resonators are made from superconducting materials with
different electron-phonon interaction strengths, tantalum (strong
interaction) and aluminium (weak interaction). The tantalum film,
150 nm thick, is sputtered on a high resistivity silicon substrate.
A 6 nm thick niobium seed layer is used to promote the growth of the
desired tantalum alpha phase \cite{face}. The critical temperature
$T_c$ is 4.43 K, the low temperature resistivity $\rho$ is 8.4 $\mu
\Omega$cm and the residual resistance ratio ($RRR$) is 3.0. A 100 nm
thick aluminium film is sputtered on silicon ($T_c$=1.25, $\rho$=1.3
$\mu \Omega$cm, $RRR$=3.7). Alternatively, a film of 250 nm thick is
sputtered on silicon ($T_c$=1.22, $\rho$=1.0 $\mu \Omega$cm,
$RRR$=6.9) and another one of 250 nm is sputtered on A-plane
sapphire ($T_c$=1.20, $\rho$=0.25 $\mu \Omega$cm, $RRR$=11). The
samples are patterned using optical lithography, followed by wet
etching for aluminium and reactive ion etching for tantalum. For
both materials quality factors in the order of $10^6$ are reached.
The sample is cooled in a cryostat with an adiabatic demagnetization
refrigerator. The sample space is surrounded by a cryoperm and a
superconducting magnetic shield. Alternatively, the sample is cooled
in a cryostat with a $^3$He sorption cooler without magnetic
shields. A GaAsP LED (1.9 eV) acts as photon source, fibre-optically
coupled to the sample box.

The complex transmission $S_{21}$ of the circuit is measured by
sweeping the frequency of the signal applied along the feedline
(Fig. \ref{figure:fig1}a). Near the resonance frequency $\omega_0$
the feedline transmission exhibits a decrease in magnitude and
traces a circle in the complex plane (full lines in Fig.
\ref{figure:fig1}b). A non-equilibrium state results in a resonance
frequency shift and broadening of the dip, and a reduction and shift
of the resonance circle in the complex plane (dashed lines in Fig.
\ref{figure:fig1}b). The actual signals (filled dot and open circle
in Fig. \ref{figure:fig1}b) are obtained by sending a continuous
wave at the equilibrium resonance frequency $\omega_0$ through the
feedline, which is amplified and mixed with a copy of the original
signal in a quadrature mixer, whose output gives the real and
imaginary part of the feedline transmission (upper inset Fig.
\ref{figure:fig1}c). The non-equilibrium response (open circle),
compared to the equilibrium response (filled dot), is characterized
by a changed phase $\theta$ and amplitude $A$, referred to a shifted
origin in the complex plane (from the equilibrium position $x_c$).

The phase $\theta$ with respect to the resonance circle center $x_c$
is given by $\theta=\arctan
[\mathrm{Im}(S_{21})/(x_c-\mathrm{Re}(S_{21}))] $ and is related to
the change in resonance frequency by: $\theta=-4Q \frac{\delta
\omega_0}{\omega_0}$, with $Q$ the resonator loaded quality factor
\cite{day}. A related change in $L_k$ is given by $\delta
\omega_0/\omega_0=-\frac{\alpha}{2} \delta L_k / L_k $, with
$\alpha$ the ratio of the kinetic to the total inductance. The phase
$\theta$ is therefore a direct measure of the change in complex
conductivity (given in the dirty limit by):
\begin{equation}
\label{equation:theta} \theta= - 2 \alpha Q \frac{\delta \sigma_2
}{\sigma_2} \Big( f(E),\Delta \Big),
\end{equation}
with $f(E)$ the electronic distribution function characterizing the
non-equilibrium and $\Delta$ the superconductor energy gap.

The amplitude $A$ depends predominantly on $\sigma_1$ and to a
smaller degree on $\sigma_2$. The amplitude is determined by the
complex transmission $S_{21}$ by:
$A=\sqrt{[\mathrm{Re}(S_{21})-x_c]^2+\mathrm{Im}(S_{21})^2}/(1-x_c)$.
On resonance $S_{21}=Q_c/(Q_c+Q_u)$ with $Q_u \propto
\sigma_2/\sigma_1$ the unloaded resonator quality factor and $Q_c$
the coupling quality factor, leading to
\begin{equation}
\label{equation:amplitude} A = 1 - 2 \frac{Q}{Q_u} \Big[
\frac{\delta \sigma_1}{\sigma_1} \Big(f(E),\Delta \Big) -
\frac{\delta \sigma_2}{\sigma_2} \Big(f(E),\Delta \Big) \Big].
\end{equation}
By measuring $A$ and $\theta$ in the frequency- and time-domain we
obtain direct information on the relaxation through the complex
conductivity of the superconducting films.

A typical pulse response is shown in Fig. \ref{figure:fig1}c. The
initial rise of the phase $\theta$ is due to the response time of
the resonator. The relaxation shows up as an exponential decay. The
right inset of Fig. \ref{figure:fig1}c shows the evolution of the
response in the transformed polar plane. These data are interpreted
as governed by one relaxation time. This is justified by performing
measurements of the noise spectrum and applying the analysis by
Wilson \textit{et al.} \cite{wilson2001}. Since the superconducting
condensate and the quasiparticle excitations form a two-level system
a Lorentzian spectrum is expected, with the relaxation time
determining the roll-off frequency. If more dominant relaxation
processes are present, the noise spectrum is no longer a single
Lorentzian \cite{wilson2004}. We have studied the superconducting
films under exposure to a \emph{continuous} photon flux. Our films
are exposed to an optical white noise signal due to photon shot
noise, resulting in fluctuations in $f(E)$. Where a single time
$\tau$ determines the relaxation process the phase or amplitude
noise spectrum is
\begin{equation}
\label{equation:stheta} S_{\theta,A} = \frac{2\hbar \Omega}{P}
\frac{r_{\theta,A}}{1+ (2\pi f \tau)^2},
\end{equation}
with $P$ the absorbed power, $\hbar \Omega$ the photon energy, and
$r_{\theta,A}$ denoting the responsivity of the phase or amplitude
to an optical signal.

\begin{figure}[!t]
    \centering
    \includegraphics[width=1\linewidth]{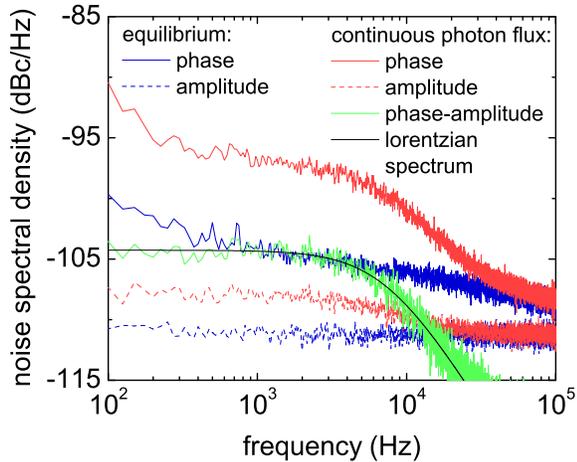}
    \caption{(Color online) The power spectral density of phase (solid line) and amplitude (dashed) in equilibrium (blue) and under a continuous photon flux (red) at a bath temperature of 310 mK.
    The cross-power spectral density (solid green) under a continuous photon flux follows a single pole Lorentzian spectrum, $S\propto [1+(2\pi f \tau)^2]^{-1}$,
    with a characteristic time of 21.7$\pm$0.3 $\mu s$ (solid black). The response time of the resonator is 0.5 $\mu s$.}
    \label{figure:noise}
\end{figure}

The measured noise power spectra of the amplitude and phase of a
tantalum sample are shown in Fig. \ref{figure:noise}. In equilibrium
the amplitude noise spectrum (dashed blue line) is flat over the
full range, and the phase noise (solid blue line) follows $1/f^a$
with $a \approx 0.25$. The amplitude noise is due to the amplifier,
remaining unchanged at frequencies far away from $\omega_0$ while
the phase noise is dominated by resonator noise \cite{gao,barends},
rolling off at a frequency corresponding to the resonator response
time (0.5 $\mu$s). Under a continuous photon flux we observe excess
noise in both amplitude (dashed red line) and phase (solid red line)
that rolls off to the equilibrium value around 8 kHz.

The difference in noise levels is equal to the difference in
responsivity: $r_A/r_\theta = 0.23$ (-13~dB), measured for this
sample. In addition, we estimate, based on 20 pW optical power
absorbed by the resonator, a phase noise level of $-94$~dBc/Hz due
to photon shot noise, which is close to the observed value. Thus we
conclude that the excess noise is due to variations in $f(E)$
induced by the photon flux. In order to eliminate the system and
resonator noise we calculate the phase-amplitude cross-power
spectral density (solid green line). We find that its spectrum is
real, indicating that variations in $f(E)$ appear as fluctuations in
the amplitude and phase without relative time delay, and that the
data follow a Lorentzian spectrum with a single time. The time
measured in the pulse response (23.0$\pm$0.5 $\mu$s) agrees with the
one determined from the noise spectrum (21.7$\pm$0.3 $\mu$s). We
have checked at several bath temperatures and found, also for
aluminium samples, only a single time. We conclude that the
relaxation time is the single dominant time in the recovery of
equilibrium.

\begin{figure}[!b]
    \centering
    \includegraphics[width=1\linewidth]{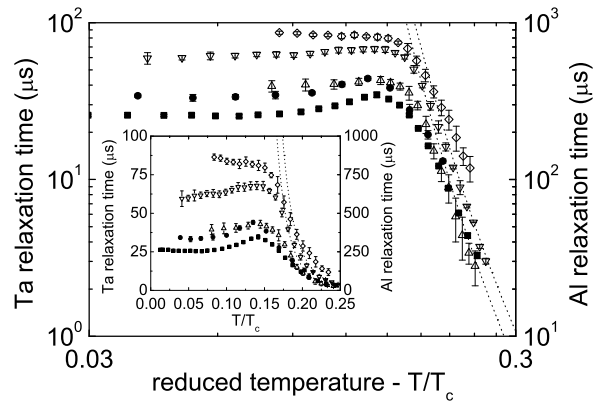}
    \caption{The relaxation times as a function of reduced bath temperature
    for 150 nm Ta on Si ($\blacksquare$, \ding{108}), 100 nm Al on Si ($\vartriangle$),
    250 nm Al on Si ($\triangledown$) and 250 nm Al on sapphire ($\Diamond$) samples.
    The inset shows the same data on a linear scale. The dotted lines are fits to the data using Eq. \ref{equation:taurec}.}
    \label{figure:tau}
\end{figure}

The measured relaxation times for temperatures down to 50 mK are
displayed in Fig. \ref{figure:tau}. The data shown are
representative for the relaxation times found in all samples of
different films. In the high temperature regime ($T/T_c \gtrsim
0.175$) the relaxation times increase for decreasing bath
temperature in a similar manner for both tantalum and aluminium
samples until a new regime is entered around $T/T_c \sim 0.15$. The
tantalum samples clearly show a non-monotonic temperature
dependence, exhibiting a maximum near $T/T_c \sim 0.15$. Two
aluminium films show a less pronounced non-monotonic temperature
dependence. We do not see a non-monotonic temperature dependence in
samples of aluminium with the lowest level of disorder (highest
$RRR$). Below $T/T_c \sim 0.1$ the relaxation times become
temperature independent at a plateau value of 25-35 $\mu$s for Ta,
390 $\mu$s for 100 nm thick Al on Si, 600 $\mu$s for 250 nm thick Al
on Si and 860 $\mu$s for 250 nm thick Al on sapphire.

The relaxation times for aluminium are measured in half wavelength
resonators where the central line is isolated from the ground plane.
For the directly connected quarter wavelength resonators a length
dependence was found. For tantalum the values are found to be length
independent in both cases. Consequently, the data shown are not
influenced by quasiparticle outdiffusion. Also, the relaxation times
remain unchanged when instead of an optical pulse a microwave pulse
at frequency $\omega_0$ is used. In this method only quasiparticle
excitations near the gap energy are created by the pair-breaking
current. This observation leads us to believe that the observed
decay is due to recombination of quasiparticles with energies near
the gap.

The exponential temperature dependence for $T/T_c \gtrsim 0.175$ is
consistent with the theory of recombination by electron-phonon
interaction \cite{kaplan}. The dotted lines in Fig. \ref{figure:tau}
follow the expression for the recombination time,
\begin{equation}
\label{equation:taurec} \frac{1}{\tau_{rec}} = \frac{1}{\tau_0}
\sqrt{\pi} \Big(\frac{2\Delta}{k T_c} \Big)^{5/2}
\sqrt{\frac{T}{T_c}} e^{-\frac{\Delta}{k T}},
\end{equation}
with $\tau_0$ a material-specific electron-phonon scattering time.
We find for 150 nm Ta on Si $\tau_0 = 42 \pm 2$ ns and for 250 nm Al
on Si $\tau_0=687 \pm 6$ ns. The deviation from the exponential rise
and the low temperature behavior is incompatible with the
established theory for electron-phonon relaxation. We assume that an
additional relaxation channel \cite{reizer} is dominant at low
temperatures, where the electron-phonon mechanism becomes too slow.

In previous experiments using superconducting tunnel junctions a
similar saturation in the quasiparticle loss has been reported. For
photon detectors inverse loss rates in the order of tens of
microseconds have been found for tantalum
\cite{verhoeve,nussbaumer,li,mazin} and hundreds of microseconds for
aluminium \cite{day}. Some of these experiments also indicated a
non-monotonic temperature dependence \cite{kozorezov2001}. Most of
these observations have been attributed to trapping states at
surfaces or in dielectrics. The fact that our similar experimental
results occur in simple superconducting films and two different
materials suggests that processes in the superconducting film itself
lead to the observed low temperature behavior.

The observed saturation in the relaxation times in our samples is
reminiscent of experiments in normal metals on inelastic scattering
in non-thermal distributions and on dephasing in weak localization
studies. The apparent saturation of the dephasing time and the
strong quasiparticle energy exchange at low temperatures have been
shown to be caused by dilute concentrations of magnetic impurities
\cite{pierre,anthore,huard,saminadayar}. It is known that in
superconductors a large density of magnetic impurities decreases the
critical temperature. For dilute magnetic impurities the local
properties are most important. In experiments with magnetic adatoms
impurity bound excitations arise \cite{yazdani}, tails in the
density of states within the gap might form and the formation of an
intragap band with growing impurity concentration are predicted
\cite{silva,balatsky}. In ongoing experiments we observe a gradual
decrease of the relaxation time with an increasing ion-implanted
magnetic impurity concentration (0-100 ppm). However, disorder plays
a role as well and further experiments are needed to clarify
possible relaxation processes \cite{kozorezovarxiv}.

In conclusion, we find that the quasiparticle relaxation times,
probed by means of the complex conductivity, saturate for both
tantalum and aluminium, below a tenth of the critical temperature.
We suggest that the saturation of the relaxation time is due to the
presence of a relaxation channel, which is not caused by the
conventional process dominated by electron-phonon interaction.

\begin{acknowledgments}
The authors thank Y. J. Y. Lankwarden for fabrication of the
devices, A. G. Kozorezov, A. A. Golubov and R. A. Hijmering for
helpful discussions and H. F. C. Hoevers for support. The work was
supported by RadioNet (EU) under contract no. RII3-CT-2003-505818
and the Netherlands Organisation for Scientific Research (NWO).
\end{acknowledgments}

\end{document}